\documentclass[12pt]{article}

\def\Journal#1#2#3#4{{\it #1} {\bf #2}, #3 (#4)}

\newcommand{\be}{\begin{equation}}
 \newcommand{\ee}{\end{equation}}

 \def\la{\mathrel{\mathpalette\fun <}}
\def\ga{\mathrel{\mathpalette\fun >}}
\def\fun#1#2{\lower3.6pt\vbox{\baselineskip0pt\lineskip.9pt
\ialign{$\mathsurround=0pt#1\hfil ##\hfil$\crcr#2\crcr\sim\crcr}}}

\newcommand{\lan}{\langle}
\newcommand{\ran}{\rangle}
\newcommand{\veA}{\mbox{\boldmath${\rm A}$}}
\newcommand{\vek}{\mbox{\boldmath${\rm k}$}}

\begin{document}
\begin{center}

{\bf CROSSOVER, FLUCTUATIONS AND ANDERSON TRANSITION IN QUARK MATTER
FORMATION } \\
\medskip
B. KERBIKOV\footnote{Talk at ICHEP'06, Moscow, 26 July - 2 August
2006}
\\
\medskip
{\it Institute of Theoretical
and Experimental Physics, \\
Moscow, Russia; email: borisk@itep.ru}
\end{center}



\begin{abstract}
We argue that there is a unique transition state of moderate density
between the nuclear matter and superconducting quark matter
alternatives. The distinguishing features of this state are
discussed.
\end{abstract}


During the last years, the investigation of quark matter at finite
temperature and density has become one of the QCD focal points.
Based on model calculations it has been predicted that at
densities 3-5 times larger than the normal nuclear density the
system is unstable with respect to the formation of quark-quark
Cooper-pair condensate. This phenomenon is called  color
superconductivity since diquarks  belong to $\bar 3$ color
channel. At present  there is a fair understanding of color
superconductivity physics in the regime of ultra-high density when
$\alpha_s$ is small. It took quite some time before it has been
realized \cite{1,2} that halfway between the nuclear matter phase
and color superconducting state lies a new state unlike anything
seen before.

Shortly after the formulation of the BCS theory it has been realized
that in one extreme case fermions can pair strongly forming compact
nuclear-like states while at the other end of the spectrum they can
pair weakly forming Cooper pairs. The continuous transition between
the two  regimes is called the BEC-BCS crossover (transition from
Bose-Einstein condensation to Bardeen-Cooper-Schrieffer regime). The
dimensionless crossover parameter in $n^{1/3}\xi$, where $n$ is the
quark number density and $\xi$ is the characteristic length of pair
correlation when the system is in the BCS regime and $\xi$ is the
root of the mean square radius of the bound state when  the system
is in the strong coupling regime. The crossover occurs at
$n^{1/3}\xi\sim 1$. Calculations within the framework of the
NJL-type models show that the gap equation acquires a nontrivial
solution $\Delta \neq 0$ when the quark chemical potential reaches
the value $\mu\simeq 0.4$ GeV which corresponds to $n^{1/3} \simeq
1$ fm$^{-1}$.  The pair size in this region is $\xi\simeq 2$ fm, so
that $n^{1/3} \xi \sim 1$ (for electron superconductors $n^{1/3}
\xi\ga 10^3$). We see that $NM\to QM$ transition brings the system
to the crossover regime and not to color superconducting state.

The transition region is also characterized by strong fluctuations
of both quark and gluon fields. The two phenomena -- crossover and
strong fluctuations, are  interrelated as we shall see from Eq.
(\ref{1}) below. Due to fluctuating quark pairs the energy
spectrum develops a pseudogap instead of the true gap typical for
the BCS regime. Fluctuation contribution to the physical
quantities and the width of the fluctuation region is measured by
the Ginzburg-Levanyuk number $Gi$ which can be estimated as \be
Gi\simeq 80 \left( \frac{T_c}{\mu}\right)^4 \simeq
\frac{0.05}{(n^{1/3}\xi)^4},\label{1}\ee where $T_c\simeq 40-50$
MeV is the critical temperature. In the transition region $T_c/\mu
\simeq 0.1$, the  crossover parameter $n^{1/3} \xi \simeq 1-2$, so
that $Gi \simeq 10^{-2}$ which is a huge number as compared to
$Gi\simeq 10^{-12} -10^{-14}$ for metal superconductors and
several magnitudes larger than for HTSC.

Fluctuating pre-formed quark pairs are responsible for the
pseudogap phenomena, while fluctuating gluon  field gives rise to
induced color diamagnetism. Color field fluctuations may be
characterized by the average $\lan \veA^2\ran$ of the gauge field
in a fixed configuration of the quark field \cite{3,4}. The
critical temperature gets lower according to the following
equation \be T'_c=T_c (1-g^2\xi^2\lan \veA^2\ran),\label{2}\ee
where $g^2= 4\pi \alpha_S$ is the strong coupling constant.

The third basic feature of the transition region is the deformation
of the  gradient term in the Ginzburg-Landau functional. In momentum
space the $G-L$ functional reads \be F_s -F_n =a |\Delta (\vek) |^2
+\frac{b}{2} |\Delta (\vek)|^4 +c\vek^2|
\Delta(\vek)|^2.\label{3}\ee The standard expression for $c$ is
$c=\nu_F \xi^2$ with $\nu_F$ being the density of states at the
Fermi level. In the  transition region the coefficient $c$ may be
expressed via the quark dynamical diffusion coefficient $D(\omega)$
which in its turn is expressed via the correlator of the two quark
Green's functions. The characteristic frequency $\omega$ depends on
the dynamics of the gluon field in the transition region which is as
yet ill understood. One may think that $\omega$ is equal  to the
gluelump mass but this is a problem for further work. An interesting
scenario of the behavior of the coefficient $c$ involving Anderson
localization is possible. In case when the quark chemical potential
$\mu$ and the quark mean free path $l$ meet the condition $\mu l\la
1$ the  system may undergo Anderson phase transition which would
lead to the decrease of the coefficient $c$ in the $G-L$
functional(\ref{3}).

Detailed considerations of some problems touched upon in this talk
as well as a list of references can be found in our recent paper
\cite{5}.
\section*{Acknowledgments}
This  work  was partially supported by the RFBR grant 06-02-17012,
Leading Scientific Schools grant 843.2006.2 and state Contract
02.445.11.7424, 2006-112.

\end{document}